\documentclass[12pt]{article}
\begin{document}
\baselineskip=0.7cm
\renewcommand{\theequation}{\arabic{section}.\arabic{equation}}
\renewcommand{\thesection}{\arabic{section}.}
\renewcommand{\thesubsection}{\arabic{section}.\arabic{subsection}}
\makeatletter
\def\section{\@startsection{section}{1}{\z@}{-3.5ex plus -1ex minus
 -.2ex}{2.3ex plus .2ex}{\large}}
\def\subsection{\@startsection{subsection}{2}{\z@}{-3.25ex plus -1ex minus
 -.2ex}{1.5ex plus .2ex}{\normalsize\it}}
\def\appendix{
\par
\setcounter{section}{0}
\setcounter{subsection}{0}
\def\thesection{\Alph{section}}}
\makeatother
\def\thefootnote{\fnsymbol{footnote}}
\begin{flushright}
hep-th/9909132\\
OU-HET 327\\
September 1999
\end{flushright}
\vspace{1cm}
\begin{center}
\Large
Derivative corrections
to Dirac-Born-Infeld Lagrangian\\
and non-commutative gauge theory

\vspace{1cm}
\normalsize
{\sc Yuji Okawa}
\footnote{
E-mail:\ \ okawa@het.phys.sci.osaka-u.ac.jp}
\\
\vspace{0.3cm}
{\it Department of Physics,
Graduate School of Science, Osaka University,\\
Toyonaka, Osaka 560-0043, Japan}

\vspace{1.3cm}
Abstract\\

\end{center}
We consider the constraints
on the effective Lagrangian of the rank-one gauge field
on D-branes
imposed by the equivalence between 
the description by ordinary gauge theory
and that by non-commutative gauge theory
in the presence of a constant $B$ field.
It is shown that
we can consistently construct
the two-derivative corrections
to the Dirac-Born-Infeld Lagrangian
up to the quartic order of field strength
and the most general form
which satisfies the constraints up to this order
is derived.

\newpage
\section{Introduction}
\setcounter{equation}{0}
There are two different descriptions
of the effective Lagrangian of the gauge fields
on D-branes
in flat space, with metric $g_{ij}$,
in the presence of  a constant Neveu-Schwarz--Neveu-Schwarz
two-form gauge field ($B$ field) $B_{ij}$.
The first one is the conventional one
in terms of ordinary gauge fields with ordinary gauge invariance.
The gauge transformation and field strength
are familiar, which are for rank-one gauge theory,
\begin{eqnarray}
\delta_\lambda A_i &=& \partial_i \lambda, \nonumber \\
F_{ij} &=& \partial_i A_j - \partial_j A_i, \nonumber \\
\delta_\lambda F_{ij} &=& 0.
\end{eqnarray}
In this formulation, the $B$-dependence
of the effective Lagrangian ${\cal L}$ 
is only in the combination $B+F$.

The other one is in terms of non-commutative gauge fields
\cite{CDS}
where the algebra of functions is deformed to
a non-commutative, associative one defined by
\begin{eqnarray}
f(x) \ast g(x) &=& \exp \left. \left(
\frac{i}{2} \theta^{ij}
\frac{\partial}{\partial \xi^i}
\frac{\partial}{\partial \zeta^j} \right)
f(x + \xi) g(x + \zeta) \right|_{\xi=\zeta=0}
\nonumber \\
&=& fg + \frac{i}{2} \theta^{ij} \partial_i f
\partial_j g + O(\theta^2),
\label{star}
\end{eqnarray}
with
\begin{equation}
\theta^{ij} =
- (2 \pi \alpha')^2 \left(
\frac{1}{g + 2 \pi \alpha' B} B \frac{1}{g - 2 \pi \alpha' B}
\right)^{ij},
\end{equation}
and the gauge transformation and field strength
for rank-one gauge theory are
correspondingly deformed to
\begin{eqnarray}
\hat{\delta}_{\hat{\lambda}} \hat{A}_i
&=& \partial_i \hat{\lambda} + i \hat{\lambda} \ast \hat{A}_i
- i \hat{A}_i \ast \hat{\lambda}, \nonumber \\
\hat{F}_{ij} &=& \partial_i \hat{A}_j - \partial_j \hat{A}_i
-i \hat{A}_i \ast \hat{A}_j +i \hat{A}_j \ast \hat{A}_i,
\nonumber \\
\hat{\delta}_{\hat{\lambda}} \hat{F}_{ij}
&=& i \hat{\lambda} \ast \hat{F}_{ij}
- i \hat{F}_{ij} \ast \hat{\lambda}.
\end{eqnarray}
The effective Lagrangian $\hat{{\cal L}} (\hat{F})$
in the latter formulation takes the same form
as the one ${\cal L} (F)$ in the former \cite{SW} except that
the product of functions is replaced
with the $\ast$ product (\ref{star})
and that Lorentz indices are contracted by the metric $G_{ij}$
which is different from the metric
$g_{ij}$ used in the description in terms of
ordinary gauge theory:
\begin{eqnarray}
G_{ij} &=& g_{ij} - (2 \pi \alpha')^2 (B g^{-1} B)_{ij}, \\
(G^{-1})^{ij} &=& \left(
\frac{1}{g + 2 \pi \alpha' B} g \frac{1}{g - 2 \pi \alpha' B}
\right)^{ij}.
\end{eqnarray}
The $B$-dependence in the latter formulation
is encoded in $\theta$, $G$ and the coupling constant.

The equivalence of the two descriptions
is recently discussed in detail \cite{SW}
which is realized by the transformation
between the non-commutative gauge field $\hat{A}$
and the ordinary one $A$,
\begin{equation}
\hat{A} (A) + \hat{\delta}_{\hat{\lambda}} \hat{A} (A)
= \hat{A} ( A + \delta_\lambda A ),
\label{A-hat}
\end{equation}
with infinitesimal $\lambda$ and $\hat{\lambda}(\lambda,A)$.
The two Lagrangians in terms of $A$ and $\hat{A}$
should be related as
\begin{equation}
{\cal L} ( B+F ) = \hat{{\cal L}} (\hat{F})
+ {\rm total~derivative},
\label{L}
\end{equation}
under the transformation (\ref{A-hat}).
This was verified in \cite{SW}
for the Dirac-Born-Infeld (DBI)
Lagrangian\footnote{
For a recent review of the Dirac-Born-Infeld theory
see \cite{Tseytlin} and references therein.
} in the approximation that
field strength is slowly varying.
We should note here that
the relation between $\hat{A}$ and $A$
(\ref{A-hat}) is determined
{\it independently} of the form
of the effective Lagrangian
and thus the condition (\ref{L}) imposes
constraints on the form of the Lagrangian
as is argued in \cite{SW}
that the DBI Lagrangian
is the only Lagrangian which satisfies (\ref{L})
in the approximation that
field strength is slowly varying.

The applicability of the argument that
the two formulations are equivalent
and the condition (\ref{L}) should be satisfied
is, however, more general
and not restricted to such cases.
It would be a non-trivial question
whether the condition (\ref{L}) can indeed
be satisfied
when we include derivative corrections
to the DBI Lagrangian.
If we assume that it is possible,
the next question to be raised is
to what extent the Lagrangian is constrained
by the condition (\ref{L}).
In the present paper, we consider the questions
for two-derivative corrections to the DBI Lagrangian
up to the quartic order of field strength
in rank-one gauge theory.
We derive the most general structure
which satisfies the condition (\ref{L})
up to this order.

The organization of this paper is as follows.
In Section 2, we first consider how $F^4$ terms
are determined by the requirement (\ref{L})
to illustrate our method for this simplest case.
We then extend our discussions
to two-derivative corrections in Section 3.
Section 4 is devoted to conclusions and discussions.

\section{Determination of $F^4$ terms}
\setcounter{equation}{0}

It is already argued in \cite{SW} that
the DBI Lagrangian satisfies the condition (\ref{L})
in the approximation that field strength is slowly varying
as we mentioned in Section 1
and most of the calculations in the present section
are nothing but the reorganization of those in \cite{SW}.
However, in addition to the purpose of
the illustration of our method which will be
applied to the determination
of derivative corrections in the next section,
it would be instructive to see how
$F^4$ terms are determined from 
the information on the $F^2$ term alone
without the information on the whole form of the Lagrangian
in order to extend our consideration to derivative corrections
of which we do not know the whole structure.

Let us begin with some preparations.
The equation (\ref{A-hat}) is solved
in the expansion with respect to $\theta$ \cite{SW}.
For rank-one gauge theory, it is given by\footnote{
It was pointed out in \cite{AK} that
there are ambiguities in perturbative solutions
to the equation (\ref{A-hat}) which are
related to gauge transformation
and to field redefinition.
However, it is easily verified that
our results will not be modified essentially
by the presence of such ambiguities
up to the order we are discussing.
}
\begin{eqnarray}
\hat{A}_i &=& A_i - \frac{1}{2} \theta^{kl}
A_k (\partial_l A_i + F_{li}) + O(\theta^2),
\label{A-hat-2} \\
\hat{\lambda} &=& \lambda + \frac{1}{2} \theta^{kl}
\partial_k \lambda A_l + O(\theta^2).
\end{eqnarray}
For the field strength,
it follows from the solution for $\hat{A}$ that
\begin{equation}
\hat{F}_{ij} = F_{ij} - \theta^{kl}
( F_{ik} F_{lj} + A_k \partial_l F_{ij} )
+ O(\theta^2).
\end{equation}
We need to expand $G^{-1}$, $\theta$,
the $\ast$ product (\ref{star}) and $\hat{F}_{ij}$
with respect to $\alpha'$:
\begin{eqnarray}
(G^{-1})^{ij} &=& (g^{-1})^{ij}
+ (2 \pi \alpha')^2
( g^{-1} B g^{-1} B g^{-1} )^{ij}
+ O(\alpha'^4), \\
\theta^{ij} &=& - (2 \pi \alpha')^2
( g^{-1} B g^{-1} )^{ij} + O(\alpha'^4), \\
f \ast g &=& fg -\frac{i}{2} (2 \pi \alpha')^2
( g^{-1} B g^{-1} )^{kl} \partial_k f \partial_l g
+ O(\alpha'^4), \\
\hat{F}_{ij} &=& F_{ij} + (2 \pi \alpha')^2
( g^{-1} B g^{-1} )^{kl}
( F_{ik} F_{lj} + A_k \partial_l F_{ij} )
+ O(\alpha'^4).
\end{eqnarray}
The discussions presented in this paper
do not depend on the dimension of space-time
on which the gauge theory is defined,
namely, the dimension of world-volume of the D-brane.
We only need to multiply an appropriate power of
$\alpha'$ to the Lagrangian to make the action
dimensionless.

Let us first verify that the $F^2$ term
\begin{equation}
{\cal L}(F) = {\rm Tr} ( g^{-1} F g^{-1} F ) + O(\alpha'^2)
\equiv (g^{-1})^{ij} F_{jk} (g^{-1})^{kl} F_{li}
+ O(\alpha'^2),
\label{F2}
\end{equation}
satisfies the condition (\ref{L}).
The left-hand side of (\ref{L}) is
\begin{eqnarray}
{\cal L}(B+F)
&=& {\rm Tr} ( g^{-1} (B+F) g^{-1} (B+F) ) + O(\alpha'^2)
\nonumber \\
&=& {\rm Tr} ( g^{-1} B )^2
+2 {\rm Tr} ( g^{-1} B g^{-1} F )
+ {\rm Tr} ( g^{-1} F )^2 + O(\alpha'^2)
\nonumber \\
&=& {\rm Tr} ( g^{-1} F )^2 + {\rm constant}
+ {\rm total~derivative}
+ O(\alpha'^2).
\end{eqnarray}
The non-commutative counterpart,
the right-hand side of (\ref{L}), is
\begin{equation}
\hat{{\cal L}} (\hat{F})
= {\rm Tr} ( G^{-1} \hat{F} \ast G^{-1} \hat{F} )
+ O(\alpha'^2)
= {\rm Tr} ( g^{-1} F g^{-1} F )
+ O(\alpha'^2).
\end{equation}
Thus (\ref{L}) is satisfied for (\ref{F2}).
This may seem trivial but is important:
The $F^2$ term (\ref{F2}) is qualified
as an initial term of a consistent Lagrangian
in the $\alpha'$ expansion.
Let us define the {\it initial term condition} as follows:
If $f(F)$ satisfies
\begin{equation}
f(B+F) = f(F) + {\rm total~derivative},
\label{initial}
\end{equation}
we say that $f(F)$ satisfies the initial term condition.
In this terminology, ${\rm Tr} ( g^{-1} F g^{-1} F )$
satisfies the initial term condition.

Now we go on to the order $O(\alpha'^2)$.
The non-commutative side
${\rm Tr} ( G^{-1} \hat{F} \ast G^{-1} \hat{F} )$
is evaluated as follows:
\begin{eqnarray}
{\rm Tr} ( G^{-1} \hat{F} \ast G^{-1} \hat{F} )
&=& {\rm Tr} ( G^{-1} \hat{F} G^{-1} \hat{F} ) + O(\alpha'^4)
\nonumber \\
&=& {\rm Tr} ( g^{-1} \hat{F} )^2
+2 ( 2 \pi \alpha')^2
{\rm Tr} ( g^{-1} B )^2 (g^{-1} F )^2
+ O(\alpha'^4)
\nonumber \\
&=& {\rm Tr} ( g^{-1} F )^2
\nonumber \\
&& +2 ( 2 \pi \alpha')^2
{\rm Tr} g^{-1} B (g^{-1} F )^3
-\frac{1}{2} ( 2 \pi \alpha')^2
{\rm Tr} ( g^{-1} B g^{-1} F )
{\rm Tr} ( g^{-1} F )^2
\nonumber \\
&& +2 ( 2 \pi \alpha')^2
{\rm Tr} ( g^{-1} B )^2 (g^{-1} F )^2
+ {\rm total~derivative} + O(\alpha'^4).
\label{GF2}
\end{eqnarray}
The existence of $O(B F^3)$ terms and $O(B^2 F^2)$ terms
in (\ref{GF2}) implies that the corresponding terms
must exist on the commutative side as well.
The sources for such terms are
${\rm Tr} ( g^{-1} (B+F) )^4$
and $( {\rm Tr} ( g^{-1} (B+F) )^2 )^2$
which are expanded as follows:
\begin{eqnarray}
{\rm Tr} ( g^{-1} (B+F) )^4
&=& {\rm Tr} ( g^{-1} F )^4
+ 4 {\rm Tr} g^{-1} B ( g^{-1} F )^3
\nonumber \\
&& + 4 {\rm Tr} ( g^{-1} B )^2 ( g^{-1} F )^2
+ 2 {\rm Tr} ( g^{-1} B g^{-1} F )^2
\nonumber \\
&& + {\rm ~constant} + {\rm total~derivative},
\label{BF4}
\\
( {\rm Tr} ( g^{-1} (B+F) )^2 )^2
&=& ( {\rm Tr} ( g^{-1} F )^2 )^2
+ 4 {\rm Tr} ( g^{-1} B g^{-1} F ) {\rm Tr} ( g^{-1} F )^2
\nonumber \\
&& + 2 {\rm Tr} ( g^{-1} B )^2  {\rm Tr} ( g^{-1} F )^2
+ 4 ( {\rm Tr} ( g^{-1} B g^{-1} F ) )^2
\nonumber \\
&& + {\rm ~constant} + {\rm total~derivative}.
\label{BF2BF2}
\end{eqnarray}
By comparing the $O(B F^3)$ terms in (\ref{GF2})
with those in (\ref{BF4}) and (\ref{BF2BF2}),
we can uniquely determine the structure at $O(\alpha'^2)$ as
\begin{equation}
\frac{1}{2} (2 \pi \alpha')^2
{\rm Tr} ( g^{-1} (B+F) )^4
-\frac{1}{8} (2 \pi \alpha')^2
( {\rm Tr} ( g^{-1} (B+F) )^2 )^2.
\label{BF4-BF2BF2}
\end{equation}
Now the comparison of the $O(B^2 F^2)$ terms
provides a consistency condition.
There are three missing terms in (\ref{GF2})
in comparison with (\ref{BF4-BF2BF2}).
The two of them are combined into total derivative:
\begin{eqnarray}
(2 \pi \alpha')^2 \left[
{\rm Tr} ( g^{-1} B g^{-1} F )^2
- \frac{1}{2} ( {\rm Tr} ( g^{-1} B g^{-1} F ) )^2
\right] = {\rm total~derivative}.
\end{eqnarray}
The last one
\begin{equation}
-\frac{1}{4} (2 \pi \alpha')^2
{\rm Tr} ( g^{-1} B )^2  {\rm Tr} ( g^{-1} F )^2
\label{missing3}
\end{equation}
can be taken care of by the $B$-dependence
of the coupling constant.
The Lagrangian ${\cal L}(F)$
in terms of the ordinary gauge theory
should be multiplied by $\sqrt{\det g}/g_s$
where $g_s$ is the string coupling constant
and we write the corresponding factor
on the non-commutative side as $\sqrt{\det G}/G_s$.
The new coupling constant $G_s$ can depend on $B$.
The $B$-dependence of $G_s$ was determined in \cite{SW}
by using the DBI Lagrangian.
In our point of view, we cannot use the DBI Lagrangian:
We are now determining its form.
It is determined perturbatively
by the presence of the term (\ref{missing3}) as follows:
\begin{equation}
\frac{\sqrt{\det G}}{G_s}
= \frac{\sqrt{\det g}}{g_s} \left[
1 -\frac{1}{4} (2 \pi \alpha')^2
{\rm Tr} ( g^{-1} B )^2 + O(\alpha'^4)
\right].
\label{G_s}
\end{equation}
This completes the consistency check of the $O(B^2 F^2)$ terms.

The presence of the structure (\ref{BF4-BF2BF2})
in turn requires the existence of
the corresponding structure on the non-commutative side,
which is
\begin{equation}
\frac{1}{2} (2 \pi \alpha')^2
{\rm Tr} ( G^{-1} \hat{F} )^4_{\rm arbitrary}
-\frac{1}{8} (2 \pi \alpha')^2
( {\rm Tr} ( G^{-1} \hat{F} )^2 )^2_{\rm arbitrary},
\end{equation}
where the subscripts ``arbitrary'' imply that
the ordering of the four field strengths in each term is arbitrary.
Since the product in the non-commutative gauge theory is
the non-commutative $\ast$ product,
we have to specify the ordering of field strengths
as in the case of $F^4$ terms in the Yang-Mills theory.
However,
the non-commutativity becomes relevant
only at higher orders in the $\alpha'$ expansion:
\begin{eqnarray}
\frac{1}{2} (2 \pi \alpha')^2
{\rm Tr} ( G^{-1} \hat{F} )^4_{\rm arbitrary}
-\frac{1}{8} (2 \pi \alpha')^2
( {\rm Tr} ( G^{-1} \hat{F} )^2 )^2_{\rm arbitrary}
\nonumber \\
= \frac{1}{2} (2 \pi \alpha')^2
{\rm Tr} ( G^{-1} \hat{F} )^4
-\frac{1}{8} (2 \pi \alpha')^2
( {\rm Tr} ( G^{-1} \hat{F} )^2 )^2
+ O(\alpha'^4),
\end{eqnarray}
where the product in the second line is ordinary one,
so the ordering problem does not matter
at the order we are discussing.
Inversely, the consideration at the present order alone
cannot constrain the ordering.
It would be interesting to see if
the discussion at higher orders can determine
or constrain the ordering.

To summarize, we have seen that the Lagrangian
\begin{eqnarray}
{\cal L}(B+F)
= \frac{\sqrt{\det g}}{g_s} \left[
{\rm Tr} ( g^{-1} (B+F) )^2
+ \frac{1}{2} (2 \pi \alpha')^2 {\rm Tr} ( g^{-1} (B+F) )^4
\right. \nonumber \\ \left.
- \frac{1}{8} (2 \pi \alpha')^2
( {\rm Tr} ( g^{-1} (B+F) )^2 )^2
+ O(\alpha'^4) \right],
\end{eqnarray}
and its non-commutative counterpart
\begin{eqnarray}
\hat{{\cal L}}(\hat{F})
= \frac{\sqrt{\det G}}{G_s} \left[
{\rm Tr} ( G^{-1} \hat{F} \ast G^{-1} \hat{F} )
+\frac{1}{2} (2 \pi \alpha')^2
{\rm Tr} ( G^{-1} \hat{F} )^4_{\rm arbitrary}
\right. \nonumber \\ \left.
-\frac{1}{8} (2 \pi \alpha')^2
( {\rm Tr} ( G^{-1} \hat{F} )^2 )^2_{\rm arbitrary}
+ O(\alpha'^4) \right],
\end{eqnarray}
coincide up to total derivative
under the definition of $G_s$ (\ref{G_s}),
namely, the condition (\ref{L}) is satisfied.
Thus we uniquely determined the $F^4$ terms as
\begin{eqnarray}
{\cal L}(F)
= \frac{\sqrt{\det g}}{g_s} \left[
{\rm Tr} ( g^{-1} F )^2
+ \frac{1}{2} (2 \pi \alpha')^2 {\rm Tr} ( g^{-1} F )^4
\right. \nonumber \\ \left.
- \frac{1}{8} (2 \pi \alpha')^2
( {\rm Tr} ( g^{-1} F )^2 )^2
+ O(\alpha'^4) \right],
\end{eqnarray}
        from the requirement (\ref{L}) alone.
The resulting Lagrangian coincides with
the $\alpha'$ expansion of the DBI Lagrangian
if it is multiplied by
\begin{equation}
-\frac{(2 \pi \alpha')^2}
{4 (2 \pi)^p (\alpha')^{\frac{p+1}{2}}}
\end{equation}
when the dimension of the space-time is $p+1$.

We would like to make a comment here.
                From the fact that
we have determined the $F^4$ terms
uniquely from the $F^2$ term
it follows that it is impossible
to organize the $F^4$ terms so as to satisfy (\ref{L})
without the $F^2$ term.
In other words, no $F^4$ structure can satisfy
the initial term condition defined in (\ref{initial}).
We can show this explicitly
by writing the most general $F^4$ terms
and seeing if they satisfy (\ref{initial}).
It is not difficult to see that the $O(B^2 F^2)$ terms
on the left-hand side cannot be arranged
to total derivative.

\section{Determination of two-derivative corrections}
\setcounter{equation}{0}
Since we have explained our strategy
in detail in Section 2,
it would not be difficult to apply it to
the two-derivative corrections
to the DBI Lagrangian.
In the first part of this section,
we construct one of consistent Lagrangians
with two derivatives
up to the quartic order of field strength.
We then derive the most general form of the Lagrangian
up to this order in the second part.

\subsection{Construction of a consistent Lagrangian}

The two-derivative corrections to the DBI Lagrangian
can first appear at order $O(\alpha')$
compared with the $F^2$ term.
Using the integration by parts and the Bianchi identity,
any term of order $O(\alpha')$
can be transformed
to the following form up to an overall constant\footnote{
We absorbed an appropriate power of $\alpha'$
into the overall constant as well to simplify the following
expressions.
}:
\begin{equation}
{\cal L} (F) =
(g^{-1})^{nm} (g^{-1})^{ij} \partial_n F_{jk}
(g^{-1})^{kl} \partial_m F_{li}.
\label{DF2}
\end{equation}
It is possible to absorb this term
into the $F^2$ term (\ref{F2})
by field redefinition.
However, we do not know
in which definition of the gauge field
the transformation (\ref{A-hat-2})
is valid in general
so we should not make such redefinition
in determining the possible form of the Lagrangian.
It is easily seen that (\ref{DF2}) satisfies
the initial term condition (\ref{initial})
since $\partial_i (B+F)_{jk} = \partial_i F_{jk}$
for a constant $B$.

To construct the non-commutative counterpart
$\hat{{\cal L}}(\hat{F})$ of this Lagrangian,
we have to replace the derivatives in (\ref{DF2})
with covariant derivatives defined by
\begin{equation}
\hat{D}_i \hat{F}_{jk} =
\partial_i \hat{F}_{jk} -i \hat{A}_i \ast \hat{F}_{jk}
+i \hat{F}_{jk} \ast \hat{A}_i,
\end{equation}
since non-commutative gauge fields are non-commutative
even when the rank is one.
Now the non-commutative Lagrangian becomes
\begin{equation}
\hat{{\cal L}}(\hat{F}) =
(G^{-1})^{nm} (G^{-1})^{ij} \hat{D}_n \hat{F}_{jk}
\ast (G^{-1})^{kl} \hat{D}_m \hat{F}_{li}.
\label{GDF2}
\end{equation}
Since the gauge transformation of
$\hat{D}_i \hat{F}_{jk}$ is
\begin{equation}
\hat{\delta}_{\hat{\lambda}} (\hat{D}_i \hat{F}_{jk})
= i \hat{\lambda} \ast \hat{D}_i \hat{F}_{jk}
-i \hat{D}_i \hat{F}_{jk} \ast \hat{\lambda},
\end{equation}
the action made from (\ref{GDF2}) is gauge invariant.

Now let us evaluate (\ref{GDF2}) in the $\alpha'$ expansion.
The covariant derivative of field strength
$\hat{D}_n \hat{F}_{ij}$ is expanded as
\begin{eqnarray}
\hat{D}_n \hat{F}_{ij}
&=& \partial_n F_{ij}
\nonumber \\
&& + (2 \pi \alpha')^2 (g^{-1} B g^{-1})^{kl}
( \partial_n (F_{ik} F_{lj} )
+ F_{nk} \partial_l F_{ij}
+ A_k \partial_n \partial_l F_{ij} )
+ O(\alpha'^4).
\end{eqnarray}
Using this result, (\ref{GDF2}) is evaluated as follows:
\begin{eqnarray}
&& (G^{-1})^{nm} (G^{-1})^{ij} \hat{D}_n \hat{F}_{jk}
\ast (G^{-1})^{kl} \hat{D}_m \hat{F}_{li}
\nonumber \\
&=&
(G^{-1})^{nm} (G^{-1})^{ij} \hat{D}_n \hat{F}_{jk}
(G^{-1})^{kl} \hat{D}_m \hat{F}_{li} + O(\alpha'^4)
\nonumber \\
&=&
(g^{-1})^{nm} (g^{-1})^{ij} \hat{D}_n \hat{F}_{jk}
(g^{-1})^{kl} \hat{D}_m \hat{F}_{li}
\nonumber \\
&& + (2 \pi \alpha')^2 \left[
(g^{-1} B g^{-1} B g^{-1})^{nm}
(g^{-1})^{ij} \partial_n F_{jk} (g^{-1})^{kl} \partial_m F_{li}
\right. \nonumber \\
&& \left. +2 (g^{-1})^{nm} (g^{-1} B g^{-1} B g^{-1})^{ij}
\partial_n F_{jk} (g^{-1})^{kl} \partial_m F_{li}
\right]
+ O(\alpha'^4)
\nonumber \\
&=&
(g^{-1})^{nm} (g^{-1})^{ij} \partial_n F_{jk}
(g^{-1})^{kl} \partial_m F_{li}
\nonumber \\
&& + (2 \pi \alpha')^2 \left[
2 (g^{-1})^{nm} (g^{-1})^{ij} \partial_n F_{jk}
(g^{-1})^{kl} (g^{-1} B g^{-1})^{pq}
( \partial_m (F_{lp} F_{qi}) + F_{mp} \partial_q F_{li} )
\right. \nonumber \\
&& -\frac{1}{2} {\rm Tr} (g^{-1} B g^{-1} F)
(g^{-1})^{nm} (g^{-1})^{ij} (g^{-1})^{kl}
\partial_n F_{jk} \partial_m F_{li} 
\nonumber \\
&& + (g^{-1} B g^{-1} B g^{-1})^{nm}
(g^{-1})^{ij} \partial_n F_{jk} (g^{-1})^{kl} \partial_m F_{li}
\nonumber \\
&& \left. +2 (g^{-1})^{nm} (g^{-1} B g^{-1} B g^{-1})^{ij}
\partial_n F_{jk} (g^{-1})^{kl} \partial_m F_{li}
\right]
\nonumber \\
&& + {\rm ~total~derivative} + O(\alpha'^4).
\label{GDF2-2}
\end{eqnarray}
Lorentz indices in most of the expressions in what follows
are contracted with respect to the metric $g_{ij}$
so that we simplify the expressions by making
$g^{-1}$ implicit as
\begin{equation}
A_i B_i \equiv (g^{-1})^{ij} A_i B_j, \quad
\partial^2 \equiv (g^{-1})^{ij} \partial_i \partial_j,
\end{equation}
unless the other metric $G_{ij}$ is explicitly used.
With this convention, (\ref{GDF2-2}) is expressed as
\begin{eqnarray}
&& (G^{-1})^{nm} (G^{-1})^{ij} \hat{D}_n \hat{F}_{jk}
\ast (G^{-1})^{kl} \hat{D}_m \hat{F}_{li}
\nonumber \\
&=&
\partial_n F_{ij} \partial_n F_{ji}
\nonumber \\
&& + (2 \pi \alpha')^2 \left[
2 \partial_n F_{ij} \partial_n F_{jk}
( B_{kl} F_{li} + F_{kl} B_{li} )
+2 \partial_n F_{ij} F_{nk} B_{kl} \partial_l F_{ji}
\right. \nonumber \\
&& \left. -\frac{1}{2}
B_{kl} F_{lk} \partial_n F_{ij} \partial_n F_{ji}
+ B_{nk} B_{kl} \partial_n F_{ij} \partial_l F_{ji}
+2 \partial_n F_{ij} \partial_n F_{jk} B_{kl} B_{li}
\right]
\nonumber \\
&& + O(\alpha'^4) + {\rm total~derivative}.
\label{GDF2-3}
\end{eqnarray}
We can easily guess the $O(\alpha'^2)$ terms
which we have to add to (\ref{DF2})
to satisfy the condition (\ref{L}) from
the $O(B \partial^2 F^3)$ part of (\ref{GDF2-3}).
By replacing $B$ in the $O(B \partial^2 F^3)$ part
of (\ref{GDF2-3}) with $F$
and taking into account the symmetry factors,
we have the following Lagrangian:
\begin{eqnarray}
{\cal L} (F) &=&
\frac{\sqrt{\det g}}{g_s} \left[
\partial_n F_{ij} \partial_n F_{ji}
+ 2 (2 \pi \alpha')^2
\partial_n F_{ij} \partial_n F_{jk} F_{kl} F_{li}
\right. \nonumber \\
&& \left. + (2 \pi \alpha')^2
F_{nk} F_{kl} \partial_n F_{ij} \partial_l F_{ji}
-\frac{1}{4} (2 \pi \alpha')^2
F_{kl} F_{lk} \partial_n F_{ij} \partial_n F_{ji}
+ O(\alpha'^4) \right].
\label{L1}
\end{eqnarray}
By expanding ${\cal L} (B+F)$,
we can see that it generates the $O(B \partial^2 F^3)$ terms
in (\ref{GDF2-3}).
The consistency check of $O(B^2 \partial^2 F^2)$ part
can be done just as in the case of Section 2.
There is one missing term in (\ref{GDF2-3})
compared with the $O(B^2 \partial^2 F^2)$ part
of ${\cal L} (B+F)$.
Precisely the same definition of
the coupling constant $G_s$ in the non-commutative
gauge theory as (\ref{G_s})
produces the missing term.
There is again the ordering ambiguity in 
the $O(\partial^2 F^4)$ terms on the non-commutative side
$\hat{{\cal L}} (\hat{F})$
but it does not matter at the order we are considering
just as in the preceding section.

Thus we have succeeded in constructing a Lagrangian (\ref{L1})
with two derivatives up to the quartic order of field strength
which satisfies the condition (\ref{L})
under the definition of $G_s$ (\ref{G_s}).
However, the Lagrangian (\ref{L1}) may not
be the unique one which satisfies the requirement (\ref{L}).
Let us reconsider the procedure by which we obtained (\ref{L1}),
namely, replacing $B$ in the $O(B \partial^2 F^3)$ part
of (\ref{GDF2-3}) with $F$.
The resulting Lagrangian can surely produce 
$O(B \partial^2 F^3)$ part of (\ref{GDF2-3}) as we have seen.
However, it may not the unique possibility.
Take the second term
\begin{equation}
2 \partial_n F_{ij} \partial_n F_{jk}
( B_{kl} F_{li} + F_{kl} B_{li} )
\label{example}
\end{equation}
on the right-hand side of (\ref{GDF2-3}) as an example.
The term 
$\partial_n F_{ij} \partial_n F_{jk} F_{kl} F_{li}$
generates (\ref{example}) when we replace $F$ with $B+F$
but 
$\partial_n F_{ij} \partial_n (F_{jk} F_{kl} F_{li})$
also generates (\ref{example}) with extra unwanted terms.
There are possibilities that
such extra terms can be arranged to total derivative
so as to satisfy (\ref{L}).
We will consider such possibilities in the next subsection.

At any rate, the fact that
we found a consistent form of two-derivative corrections
(\ref{L1}) at least
at the current order in the $\alpha'$ expansion
is non-trivial and interesting itself.
It remains to be investigated
whether it persists to higher orders.

\subsection{Solutions
to the initial term condition}

Let us go back to the problem
whether or not the Lagrangian (\ref{L1}) is the unique one
which satisfies (\ref{L}).
Assume that there is another Lagrangian
$\tilde{{\cal L}} (F)$ satisfying (\ref{L})
which coincides with ${\cal L} (F)$ (\ref{L1})
at $O(\partial^2 F^2)$ but differs at $O(\partial^2 F^4)$.
Then the difference $\tilde{{\cal L}} (F) - {\cal L} (F)$
also satisfies the condition (\ref{L})
but it does not have $O(\partial^2 F^2)$ part.
Therefore the $O(\partial^2 F^4)$ part of it
must satisfy the initial term condition (\ref{initial}).
Now the problem of uniqueness reduced to the question
whether there are solutions of the form $O(\partial^2 F^4)$
to the initial term condition.

What we should do is now clear.
First write the most general terms of order $O(\partial^2 F^4)$
and replace $F$ with $B+F$.
The resulting terms are quadratic with respect to $B$
so that we should look for combinations of terms
such that the $O(B)$ and $O(B^2)$ parts
are arranged to total derivatives respectively.

Any term of order $O(\partial^2 F^4)$ can be
transformed to the following form
using the integration by parts
and the Bianchi identity \cite{AT1}:
\begin{equation}
{\cal L} = \sum_{i=1}^{7} b_i J_i,
\end{equation}
where
\begin{eqnarray}
&& J_1 = \partial_n F_{ij} \partial_n F_{ji} F_{kl} F_{lk},
\quad
J_2 = \partial_n F_{ij} \partial_n F_{jk} F_{kl} F_{li},
\nonumber \\
&& J_3 = F_{ni} F_{im} \partial_n F_{kl} \partial_m F_{lk},
\quad
J_4 = \partial_n F_{ni} \partial_m F_{im} F_{kl} F_{lk},
\nonumber \\
&& J_5 = -\partial_n F_{ni} \partial_m F_{ij} F_{jk} F_{km},
\quad
J_6 = \partial^2 F_{ij} F_{ji} F_{kl} F_{lk},
\nonumber \\
&& J_7 = \partial^2 F_{ij} F_{jk} F_{kl} F_{li},
\quad
\partial^2 F_{ij}
= \partial_i \partial_k F_{kj} - \partial_j \partial_k F_{ki}.
\label{AT}
\end{eqnarray}
We will call this basis $\{ J_i \}$
as the Andreev-Tseytlin basis.
This basis is useful when we consider
field redefinition
because the first three coefficients $b_1$, $b_2$ and $b_3$
in this basis do not change
under field redefinition and unambiguous \cite{AT2}.
However, the following basis will turn out to be
more convenient for the problem at hand:
\begin{equation}
{\cal L} = \sum_{i=1}^{5} a_i J_i
+ a_6 J'_6 + a_7 J'_7,
\end{equation}
where
\begin{equation}
J'_6 = F_{ij} \partial_n F_{ji} F_{kl} \partial_n F_{lk},
\quad
J'_7 = F_{ij} \partial_n F_{jk} F_{kl} \partial_n F_{li}.
\end{equation}
The two bases are related as follows:
\begin{eqnarray}
&& J'_6 = - \frac{1}{2} J_1 - \frac{1}{2} J_6,  \quad
J'_7 = -2 J_2 - J_7,
\nonumber \\
&& J_6 = - J_1 -2 J'_6, \quad
J_7 = -2 J_2 - J'_7.
\end{eqnarray}
Let us denote the $O(B^n)$ part of $J_i$
with its $F$ replaced by $B+F$ as $J_i (B^n)$
(and similarly for $J'_i$).

First consider the $O(B^2)$ part.
Explicit expressions for $J_i (B^2)$ and $J'_i (B^2)$,
and their variations with respect to $A_i$ are
\begin{eqnarray}
&& J_1 (B^2) = \partial_n F_{ij} \partial_n F_{ji} B_{kl} B_{lk},
\quad
\delta J_1 (B^2) = 4 B_{kl} B_{lk} \delta A_i
\partial^2 \partial_j F_{ij},
\nonumber \\
&& J_2 (B^2) = \partial_n F_{ij} \partial_n F_{jk} B_{kl} B_{li},
\nonumber \\
&& \delta J_2 (B^2) = 2 B_{km} B_{ml} \delta A_i
\partial^2 \partial_l F_{ik}
-2 B_{im} B_{mj} \delta A_i
\partial^2 \partial_k F_{kj},
\nonumber \\
&& J_3 (B^2) = B_{ni} B_{im} \partial_n F_{kl} \partial_m F_{lk},
\quad
\delta J_3 (B^2) = 4 B_{km} B_{ml} \delta A_i
\partial_k \partial_j \partial_l F_{ij},
\nonumber \\
&& J_4 (B^2) = \partial_n F_{ni} \partial_m F_{im} B_{kl} B_{lk},
\quad
\delta J_4 (B^2) = 2 B_{kl} B_{lk} \delta A_i
\partial^2 \partial_j F_{ij},
\nonumber \\
&& J_5 (B^2) = -\partial_n F_{ni} \partial_m F_{ij} B_{jk} B_{km},
\quad
\delta J_5 (B^2) = 2 B_{km} B_{ml} \delta A_i
\partial_j \partial_k \partial_l F_{ji},
\nonumber \\
&& J'_6 (B^2) = B_{ij} \partial_n F_{ji} B_{kl} \partial_n F_{lk},
\quad
\delta J'_6 (B^2) = 4 B_{ij} B_{kl} \delta A_i
\partial^2 \partial_j F_{lk},
\nonumber \\
&& J'_7 (B^2) = B_{ij} \partial_n F_{jk} B_{kl} \partial_n F_{li},
\quad
\delta J'_7 (B^2) = 4 B_{ij} B_{kl} \delta A_i
\partial^2 \partial_l F_{jk},
\end{eqnarray}
where total derivatives are neglected.
We can see the four different structures 
$B_{kl} B_{lk} \delta A_i$,
$B_{im} B_{mj} \delta A_i$,
$B_{km} B_{ml} \delta A_i$
and $B_{ij} B_{kl} \delta A_i$
in $\delta J_i (B^2)$ and $\delta J'_i (B^2)$.
They must vanish respectively
for a combination of $J_i (B^2)$ and $J'_i (B^2)$
to be total derivative.
Solving this condition, we found that
the most general combinations of $J_i (B^2)$ and $J'_i (B^2)$
which are total derivatives are
\begin{equation}
{\cal L} = a_1 ( J_1 - 2 J_4 )
+ a_3 ( J_3 + 2 J_5 )
+ a_6 ( J'_6 -2 J'_7 ).
\label{necessary-1}
\end{equation}
This is a necessary condition for the initial term condition
(\ref{initial}).
It follows that the number of independent solutions
to the initial term condition is three at most.

Next consider the $O(B)$ part.
$J_i (B)$ and $J'_i (B)$ are
\begin{eqnarray}
&& J_1 (B) = 2 \partial_n F_{ij} \partial_n F_{ji} B_{kl} F_{lk},
\quad
J_4 (B) = 2 \partial_n F_{ni} \partial_m F_{im} B_{kl} F_{lk},
\nonumber \\
&& J_3 (B) = 2 B_{ni} F_{im} \partial_n F_{kl} \partial_m F_{lk},
\quad
J_5 (B) = -\partial_n F_{ni} \partial_m F_{ij} B_{jk} F_{km}
-\partial_n F_{ni} \partial_m F_{ij} F_{jk} B_{km},
\nonumber \\
&& J'_6 (B) = 2 B_{ij} \partial_n F_{ji} F_{kl} \partial_n F_{lk},
\quad
J'_7 (B) = 2 B_{ij} \partial_n F_{jk} F_{kl} \partial_n F_{li},
\end{eqnarray}
where we omitted $J_2 (B)$ since its appearance
in solutions is forbidden by the necessary condition
(\ref{necessary-1}).
In this case, we can divide
$\delta J_i (B)$ and $\delta J'_i (B)$
into two parts: terms with $B_{ij} \delta A_i$
and terms with $B_{kl} \delta A_i$.
Both parts must vanish respectively
for a combination of
$J_i (B)$ and $J'_i (B)$
to be total derivative.

Consider the terms with $B_{ij} \delta A_i$.
The relevant combinations in (\ref{necessary-1}) are
\begin{eqnarray}
\delta [J_1 (B) -2 J_4 (B)]
&=& 8 B_{ij} \delta A_i
\partial_j \partial_n
(F_{nl} \partial_m F_{lm} - \partial_m F_{nl} F_{lm})
+ {\rm terms~with~} B_{kl} \delta A_i,
\nonumber \\
\delta [J_3 (B) +2 J_5 (B)]
&=& 2 B_{ij} \delta A_i
\partial_n (-\partial_j F_{kl} \partial_n F_{lk}
+2 \partial_l F_{lk} \partial_n F_{kj})
+ {\rm terms~with~} B_{kl} \delta A_i,
\nonumber \\
\delta [J'_6 (B) -2 J'_7 (B)]
&=& 4 B_{ij} \delta A_i
\partial_n (\partial_j F_{kl} \partial_n F_{lk}
-2 \partial_l F_{lk} \partial_n F_{kj})
+ {\rm terms~with~} B_{kl} \delta A_i,
\nonumber \\
\end{eqnarray}
where total derivatives are neglected.
We found that the only combination
where the terms with $B_{ij} \delta A_i$ vanish is\footnote
{
We can show that the $B_{ij} \delta A_i$ part
of $\delta [J_1 (B) -2 J_4 (B)]$
is not proportional to that of $\delta [J_3 (B) +2 J_5 (B)]$
(or equivalently to that of $\delta [J'_6 (B) -2 J'_7 (B)]$)
as follows.
The latter contains the structure $B_{ij} \delta A_i A_j$
while it is absent from the former.
Moreover, both of them have the structure
$B \delta A \partial^3 A \partial^2 A$
but the index $j$ in each term of the former
belongs to one of the derivatives in $\partial^3 A$
while it is not the case for the latter.
}
\begin{equation}
\delta \left[( J_3 (B) +2 J_5 (B) )
+ \frac{1}{2} ( J'_6 (B) -2 J'_7 (B) ) \right]
= {\rm terms~with~} B_{kl} \delta A_i.
\label{variation}
\end{equation}
Thus we have obtained an additional necessary condition
for the initial term condition (\ref{initial}),
which is
\begin{equation}
{\cal L} = a_3 \left[ ( J_3 + 2 J_5 )
+ \frac{1}{2} ( J'_6 -2 J'_7 ) \right].
\label{necessary-2}
\end{equation}
Now the number of independent solutions reduced to one at most.
In order to determine if the particular combination
(\ref{necessary-2}) satisfies
the initial term condition (\ref{initial}),
it remains to evaluate the $B_{kl} \delta A_i$ part
of (\ref{variation}) to see if it vanishes.
It would be straightforward to do that
but it requires rather tedious calculations.
We will take a different approach in the following
instead of performing such explicit evaluation.

Let us repeat the calculations in the previous subsection
with (\ref{DF2}) replaced by
\begin{equation}
{\cal L} (F) = \partial_i F_{ik} \partial_j F_{jk},
\end{equation}
although this Lagrangian is proportional to (\ref{DF2})
up to total derivative as we mentioned before.
The non-commutative counterpart of this Lagrangian
$(G^{-1})^{ij} (G^{-1})^{kl} (G^{-1})^{nm}
\hat{D}_i \hat{F}_{jn} \ast \hat{D}_k \hat{F}_{ln}$
is expanded with respect to $\alpha'$ as follows:
\begin{eqnarray}
&& (G^{-1})^{ij} (G^{-1})^{kl} (G^{-1})^{nm}
\hat{D}_i \hat{F}_{jn} \ast \hat{D}_k \hat{F}_{lm}
\nonumber \\
&=&
\partial_i F_{ik} \partial_j F_{jk}
\nonumber \\
&& + (2 \pi \alpha')^2 \left[
2 B_{kl} \partial_i ( F_{ik} F_{lj} ) \partial_n F_{nj}
+2 B_{kl} F_{ik} \partial_l F_{ij} \partial_n F_{nj}
\right. \nonumber \\
&& \left. -\frac{1}{2}
B_{kl} F_{lk} \partial_i F_{ij} \partial_n F_{nj}
+2 B_{im} B_{ml} \partial_i F_{lj} \partial_k F_{kj} 
+ B_{jm} B_{mn} \partial_i F_{ij} \partial_k F_{kn}
\right]
\nonumber \\
&& + O(\alpha'^4) + {\rm total~derivative}.
\label{GDF2-2-2}
\end{eqnarray}
As we have done in the previous case,
it is not difficult to find the following Lagrangian
\begin{eqnarray}
{\cal L} (F) &=&
\frac{\sqrt{\det g}}{g_s} \left[
\partial_i F_{ik} \partial_j F_{jk}
+ 2 (2 \pi \alpha')^2
F_{im} F_{ml} \partial_i F_{lj} \partial_k F_{kj}
\right. \nonumber \\
&& \left. + (2 \pi \alpha')^2
F_{jm} F_{mn} \partial_i F_{ij} \partial_k F_{kn}
-\frac{1}{4} (2 \pi \alpha')^2
F_{kl} F_{lk} \partial_i F_{ij} \partial_n F_{nj}
+ O(\alpha'^4) \right],
\nonumber \\
\label{L2}
\end{eqnarray}
which generates the $O(B \partial^2 F^3)$
and $O(B^2 \partial^2 F^2)$ parts of (\ref{GDF2-2-2})
under the definition of $G_s$ (\ref{G_s}).
This is another Lagrangian which satisfies
the condition (\ref{L}).

The $O(\partial^2 F^4)$ terms of both Lagrangians
(\ref{L1}) and (\ref{L2}) can be expressed
in the Andreev-Tseytlin basis (\ref{AT}) as follows:
\begin{eqnarray}
{\cal L}_1 &=&
\frac{\sqrt{\det g}}{g_s} \left[
\partial_n F_{ij} \partial_n F_{ji}
+ (2 \pi \alpha')^2
\left( 2 J_2 + J_3 -\frac{1}{4} J_1 \right)
+ O(\alpha'^4)
\right],
\\
{\cal L}_2 &=&
\frac{\sqrt{\det g}}{g_s} \left[
\partial_i F_{ik} \partial_j F_{jk}
+ (2 \pi \alpha')^2
\left( J_5 - \frac{1}{8} J_6 + \frac{1}{2} J_7 \right)
\right. \nonumber \\
&& \qquad \qquad \left. + {\rm ~total~derivative}
+ O(\alpha'^4)
\right],
\end{eqnarray}
where we have used
\begin{eqnarray}
F_{jm} F_{mn} \partial_i F_{ij} \partial_k F_{kn}
= -\frac{1}{4} J_4 - J_5
-\frac{1}{8} J_6 + \frac{1}{2} J_7
+ {\rm total~derivative}.
\end{eqnarray}
Since
\begin{equation}
\partial_n F_{ij} \partial_n F_{ji}
= -2 \partial_i F_{ik} \partial_j F_{jk}
+ {\rm total~derivative},
\label{DFDF-relation}
\end{equation}
the $O(\partial^2 F^4)$ part of
$(-2 {\cal L}_2) - {\cal L}_1$
satisfies the initial term condition (\ref{initial}),
which is
\begin{eqnarray}
&& (-2 {\cal L}_2) - {\cal L}_1
\nonumber \\
&=& \frac{(2 \pi \alpha')^2 \sqrt{\det g}}{g_s} \left[
\frac{1}{4} J_1 -2 J_2 - J_3 
-2 J_5 + \frac{1}{4} J_6 - J_7
+ {\rm total~derivative}
+ O(\alpha'^4) \right]
\nonumber \\
&=&
\frac{(2 \pi \alpha')^2 \sqrt{\det g}}{g_s} \left[
- J_3  -2 J_5 - \frac{1}{2} J'_6 + J'_7
+ {\rm total~derivative}
+ O(\alpha'^4) \right].
\end{eqnarray}
This precisely coincides with the combination
appeared in the necessary condition (\ref{necessary-2}).
Thus we have shown that
the combination of $O(\partial^2 F^4)$ terms
(\ref{necessary-2}) satisfies the initial term condition
(\ref{initial})
and that it is the only solution
since we have shown that
the number of independent solutions is one at most.

To summarize,
we derived the most general form of the Lagrangian
with two derivatives 
which satisfies (\ref{L})
up to the quartic order of field strength.
Our result is
\begin{eqnarray}
{\cal L} (F)
&=& \frac{\sqrt{\det g}}{g_s} \left[
a \partial_n F_{ij} \partial_n F_{ji}
+ a (2 \pi \alpha')^2 \left(
-\frac{1}{4} J_1 + 2 J_2 + J_3
\right)
\right. \nonumber \\
&& \left. + b (2 \pi \alpha')^2 \left(
- J_3  -2 J_5 - \frac{1}{2} J'_6 + J'_7
\right)
+ O(\alpha'^4)
\right]
\nonumber \\
&=& \frac{\sqrt{\det g}}{g_s} \left[
a \partial_n F_{ij} \partial_n F_{ji}
+ ( a-b ) (2 \pi \alpha')^2 \left(
-\frac{1}{4} J_1 + 2 J_2 + J_3
\right)
\right. \nonumber \\
&& \left. + b (2 \pi \alpha')^2 \left(
-2 J_5 + \frac{1}{4} J_6 - J_7
\right)
+ O(\alpha'^4)
\right],
\end{eqnarray}
where $a$ and $b$ are arbitrary constants.
Furthermore, the definition of $G_s$ coincides with
that for the Lagrangian without derivatives
in the previous section
so that we can superpose the two Lagrangians
without violating the condition (\ref{L}).
The resulting expression
with the appropriate overall factor of $\alpha'$
for $p+1$ space-time dimensions is
\begin{eqnarray}
&& {\cal L} (F) =
\frac{(2 \pi \alpha')^2 \sqrt{\det g}}
{g_s (2 \pi)^p (\alpha')^{\frac{p+1}{2}}} \left[
c F_{ij} F_{ji}
+ a (2 \pi \alpha') \partial_n F_{ij} \partial_n F_{ji}
\right. \nonumber \\
&& + c (2 \pi \alpha')^2 \left(
\frac{1}{2} F_{ij} F_{jk} F_{kl} F_{li}
- \frac{1}{8} F_{ij} F_{ji} F_{kl} F_{lk}
\right)
\nonumber \\
&& + ( a-b ) (2 \pi \alpha')^3 \left(
-\frac{1}{4} \partial_n F_{ij} \partial_n F_{ji} F_{kl} F_{lk}
+ 2 \partial_n F_{ij} \partial_n F_{jk} F_{kl} F_{li}
+ F_{ni} F_{im} \partial_n F_{kl} \partial_m F_{lk}
\right)
\nonumber \\ && \left.
+ b (2 \pi \alpha')^3 \left(
2 \partial_n F_{ni} \partial_m F_{ij} F_{jk} F_{km}
+ \frac{1}{4} \partial^2 F_{ij} F_{ji} F_{kl} F_{lk}
- \partial^2 F_{ij} F_{jk} F_{kl} F_{li}
\right)
+ O(\alpha'^4)
\right],
\nonumber \\
\label{final}
\end{eqnarray}
where $a$, $b$ and $c$ are arbitrary constants.
This is the most general form of the Lagrangian
which satisfies the condition (\ref{L})
up to two derivatives
and up to the quartic order of field strength\footnote
{
We can show that
there are no non-vanishing
$O(F^3)$ and $O(\partial^2 F^3)$ terms
in rank-one gauge theory.
}.

\section{Conclusions and discussions}
\setcounter{equation}{0}

We considered the constraints on
the form of the effective Lagrangian
of the rank-one gauge field on D-branes
imposed by the condition that
the two descriptions in terms of
the ordinary and non-commutative gauge theories
in the presence of a constant $B$ field
are equivalent and
are related by (\ref{L}).
We first explained
how the form of $F^4$ terms is uniquely
determined from the information on the $F^2$ term alone
by the condition (\ref{L}).
We then applied our method
to two-derivative terms
and derived the most general form of them
up to the quartic order of field strength.
The result is summarized in (\ref{final}).

Our result shows that
the equivalence of the two descriptions
can persist beyond the approximation that
the field strength is slowly varying
at least to the first non-trivial order
in the $\alpha'$ expansion.
Moreover, we found that
the requirement of the equivalence
highly constrains the form of the effective Lagrangian.
Not only the equivalence of the two descriptions
is important conceptually
but also it may be useful practically.
We hope that our approach provides
a new perspective on the analysis of the dynamics
of the gauge field on D-branes.

Finally, let us compare our result with
ones obtained from other methods
and discuss possible future direction
of our approach.
It would be helpful to discuss
how our final result (\ref{final}) behaves
under field redefinition
in comparing with results in the literature.
As we mentioned before,
the coefficients in front of
the $O(\partial^2 F^2)$ term,
$J_4$, $J_5$, $J_6$ and $J_7$
in the Andreev-Tseytlin basis
change under field redefinition.
We can make them vanish if we redefine
the gauge field $A_i$ as follows:
\begin{eqnarray}
&& \tilde{A}_i = A_i
- \frac{a}{2 c} (2 \pi \alpha') \partial_j F_{ji}
- \frac{a^2}{8 c^2} (2 \pi \alpha')^2
\partial^2 \partial_j F_{ji}
- \frac{a^3}{16 c^3} (2 \pi \alpha')^3
\partial^4 \partial_j F_{ji}
\nonumber \\
&& + (2 \pi \alpha')^3 \left(
-\frac{a - b}{8 c} \partial_n F_{in} F_{kl} F_{kl}
- \frac{a - 2b}{2 c} \partial_n F_{ik} F_{nl} F_{lk}
- \frac{a - b}{2 c} \partial_n F_{kn} F_{il} F_{lk}
\right).
\end{eqnarray}
Then the Lagrangian (\ref{final})
is rewritten in terms of
$\tilde{F}_{ij} =
\partial_i \tilde{A}_j - \partial_j \tilde{A}_i$
as
\begin{eqnarray}
&& {\cal L} =
\frac{(2 \pi \alpha')^2 \sqrt{\det g}}
{g_s (2 \pi)^p (\alpha')^{\frac{p+1}{2}}} \left[
c \tilde{F}_{ij} \tilde{F}_{ji}
+ c (2 \pi \alpha')^2 \left(
\frac{1}{2} \tilde{F}_{ij} \tilde{F}_{jk}
\tilde{F}_{kl} \tilde{F}_{li}
- \frac{1}{8} \tilde{F}_{ij} \tilde{F}_{ji}
\tilde{F}_{kl} \tilde{F}_{lk}
\right)
\right. \nonumber \\
&& + ( a-b ) (2 \pi \alpha')^3 \left(
-\frac{1}{4}
\partial_n \tilde{F}_{ij} \partial_n \tilde{F}_{ji}
\tilde{F}_{kl} \tilde{F}_{lk}
+ 2 \partial_n \tilde{F}_{ij} \partial_n \tilde{F}_{jk}
\tilde{F}_{kl} \tilde{F}_{li}
+ \tilde{F}_{ni} \tilde{F}_{im}
\partial_n \tilde{F}_{kl} \partial_m \tilde{F}_{lk}
\right)
\nonumber \\ && \left.
+ O(\alpha'^4)
\right],
\label{final-2}
\end{eqnarray}
where total derivatives are neglected.
It can be seen from this expression
that the condition (\ref{L}) determines
the coefficients which do not change
under field redefinition
almost uniquely except some overall constants.

Now let us compare (\ref{final-2}) with
results obtained from other methods.
The derivative corrections to the DBI Lagrangian
were derived from the string four-point
amplitude \cite{AT1} or from
the two-loop $\beta$-function
in the open string $\sigma$ model \cite{AT2} \footnote{
Very recently, the derivative corrections to
the D-brane action were derived from
the method of generalized boundary state
for bosonic string theory \cite{Hashimoto1}
and for superstring theory \cite{Hashimoto2}.
}.
The $O(\partial^2 F^4)$ terms
in the bosonic string case are proportional to\footnote{
This expression is slightly different from
(4) in \cite{AT2} but
the author was informed of
a misprint in (4) of \cite{AT2}:
the last coefficient $b_3$ should have sign $+$.
}
\begin{equation}
-\frac{1}{4} J_1 -2 J_2 + J_3
\label{AT-J}
\end{equation}
while our result is proportional to
\begin{equation}
-\frac{1}{4} J_1 +2 J_2 + J_3.
\label{Okawa-J}
\end{equation}
These are very close but differ in a sign.
We do not understand the origin of such discrepancy.
For the superstring case, it was found that
$O(\partial^2 F^4)$ terms vanish \cite{AT1, Hashimoto2}.
This is consistent with our result
because our method did not determine
the overall factor and allows it to vanish.
It should be clarified whether or not
the discrepancy is characteristic
of bosonic strings.

In this paper, we have concentrated on
rank-one gauge theory.
One of possible extensions of our approach
is to consider higher-rank gauge theory.
In particular, it would be an interesting question
whether our approach can constrain the ordering
of non-Abelian field strengths.
The perturbative solution to (\ref{A-hat})
for the higher-rank case is already presented
in \cite{SW} and in fact
it is not difficult to see that
the calculations presented in Section 2
can be extended to the higher-rank case as well
at the order we have considered.
However, the discussion at this order
could not determine the ordering of field strengths.
It would deserve to extend our consideration
to higher orders to discuss the problem.

Another motivation for extension to higher orders
is to investigate how terms with
different numbers of derivatives are related
by the condition (\ref{L}).
The constraints on terms without derivatives
presented in Section 2
and those on two-derivative corrections
in Section 3 are almost independent
at the order
which we have considered
except that the definitions of $G_s$
on both sides
must be the same to superpose the two Lagrangians.
However, the independence may not persist to higher orders.
This problem would be more important
for the higher-rank case where
the separation between field strengths
and covariant derivatives becomes ambiguous.

\vspace{0.5cm}
\noindent
{\it Note added}

In proving that the combination
(\ref{necessary-2}) satisfies the initial term condition,
it was assumed that the non-commutative counterpart of
the relation (\ref{DFDF-relation}) holds as well.
However, it turned out that this is not the case since from
\begin{eqnarray*}
\hat{D}_n \hat{F}_{ij} \ast \hat{D}_n \hat{F}_{ji}
&=& 2 \hat{F}_{ik} \ast \hat{D}_j \hat{D}_i \hat{F}_{jk}
+ {\rm total~derivative}, \\
-2 \hat{D}_i \hat{F}_{ik} \ast \hat{D}_j \hat{F}_{jk}
&=& 2 \hat{F}_{ik} \ast \hat{D}_i \hat{D}_j \hat{F}_{jk}
+ {\rm total~derivative},
\end{eqnarray*}
it follows that
\begin{eqnarray*}
&& \hat{D}_n \hat{F}_{ij} \ast \hat{D}_n \hat{F}_{ji}
+2 \hat{D}_i \hat{F}_{ik} \ast \hat{D}_j \hat{F}_{jk} \\
&=& 2 \hat{F}_{ik} \ast [ \hat{D}_j, \hat{D}_i ] \hat{F}_{jk}
+ {\rm total~derivative} \\
&=& -4i \hat{F}_{ij} \ast \hat{F}_{jk} \ast \hat{F}_{ki}
+ {\rm total~derivative} \\
&=& -2 (2 \pi \alpha')^2
B_{nm} F_{ij} \partial_n F_{jk} \partial_m F_{ki}
+ O(\alpha'^4)
+ {\rm total~derivative}.
\end{eqnarray*}
Thus the conclusion which we can derive from the fact that
the Lagrangians ${\cal L}_1$ and ${\cal L}_2$
satisfy the condition (\ref{L}) is not that
the combination
$$
{\cal F} (F) \equiv
-\frac{1}{4} J_1 +2 J_2 +J_3 +2 J_5 -\frac{1}{4} J_6 +J_7
$$
satisfies the initial term condition
but that
$$
{\cal F} (B+F) = {\cal F} (F)
- 2 B_{nm} F_{ij} \partial_n F_{jk} \partial_m F_{ki}
+ {\rm total~derivative},
$$
so that there is no solution of the form $O(\partial^2 F^4)$
to the initial term condition.

This does not change our final result (\ref{final}),
however the reason why we can add the part proportional to $b$
in (\ref{final}) is not that it satisfies the initial term condition
but that we can add the term
$2 \hat{F}_{ik} \ast [ \hat{D}_j, \hat{D}_i ] \hat{F}_{jk}
= -4i \hat{F}_{ij} \ast \hat{F}_{jk} \ast \hat{F}_{ki}$
which vanishes in the commutative limit
when we construct the Lagrangian on the non-commutative side.
This ambiguity in constructing $\hat{{\cal L}}(\hat{F})$ from
${\cal L}(F)$ is characteristic of the rank-one gauge theory
because the $F^3$ term no longer vanishes for higher-rank cases
and if we could succeed in generalizing the part proportional to $b$
in (\ref{final}) to the higher-rank cases,
its existence would be naturally understood
by the fact that
the $F^3$ term satisfies the initial term condition.

Furthermore, several important developments in our understanding
have been made recently \cite{OT}.
It has turned out that
it is in general possible to constrain the effective Lagrangian
without assuming the form of the field redefinition (\ref{A-hat-2})
and its form is rather regarded
as a consequence of the compatibility
of the description by non-commutative gauge theory
with that by ordinary gauge theory.
Moreover, it has turned out that
gauge-invariant but $B$-dependent corrections
to (\ref{A-hat-2}) are generally possible
and necessary for some cases
including the case of bosonic string theory,
which resolves
the discrepancy between (\ref{AT-J}) and (\ref{Okawa-J}).

\vspace{0.4cm}
\noindent
Acknowledgements

The motivation for this work was developed
during the workshop Summer Institute '99
held at Yamanashi, Japan.
The author expresses his gratitude to the organizers
for providing such opportunity
and to the participants for stimulating discussions.
In particular, the author would like to thank
K. Hashimoto
for interesting communications on
non-commutative gauge theory and
the derivative corrections to
the DBI Lagrangian,
and K. Okuyama and S. Terashima
for discussions on non-commutative gauge theory.
This work was supported in part
by the Japan Society for the Promotion of Science
under the Postdoctoral Research Program (No. 11-01732).

\end{document}